\documentclass[a4paper,twoside]{article}

\usepackage{epsfig}
\usepackage{subcaption}
\usepackage{calc}
\usepackage{amssymb}
\usepackage{amstext}
\usepackage{amsmath}
\usepackage{amsthm}
\usepackage{multicol}
\usepackage{pslatex}
\usepackage{apalike}
\usepackage{algorithm2e}
\usepackage[bottom]{footmisc}
\usepackage{SCITEPRESS}     

\usepackage{quantikz}

\begin{document}

\title{FALQON-MST: A Fully Quantum Framework for Graph Optimization in Vision Systems}

\author{
Double-blind review
}

 \author{
 \authorname{
 Guilherme E. L. Pexe\sup{1}\orcidAuthor{0000-0003-3389-9510},
 Lucas A. M. Rattighieri\sup{2}\orcidAuthor{0009-0008-9627-6357},
 Leandro A. Passos\sup{1}\orcidAuthor{0000-0003-3529-3109},
 Douglas Rodrigues\sup{1}\orcidAuthor{0000-0003-0594-3764},
 Danilo S. Jodas\sup{1}\orcidAuthor{0000-0002-0370-1211},
 Jo\~ao P. Papa\sup{1}\orcidAuthor{0000-0002-6494-7514},
 Kelton A. P. da Costa\sup{1}\orcidAuthor{0000-0001-5458-3908}
 }
 \vspace{1ex}
 \affiliation{\sup{1}S\~ao Paulo State University, School of Sciences, Av. Eng. Lu\'is Edmundo Carrijo Coube, 2085, Bauru, Brazil}
 \email{\{guilherme.pexe, leandro.passos, d.rodrigues, danilo.jodas, joao.papa, kelton.costa\}@unesp.br}
 \vspace{-1ex}
 \affiliation{\sup{2}Gleb Wataghin Institute of Physics, University of Campinas, 13083-859, Campinas-SP, Brazil}
 \email{lucas.rattighieri@dac.unicamp.br} 
 \vspace{-2ex}
 }

\keywords{Quantum Optimization, Feedback-Based Quantum Algorithms, Minimum Spanning Tree.}

\abstract{Finding the minimum spanning tree (MST) of a graph is an important task in computer vision, as it enables a sparse and low-cost representation of connectivity among elements (such as superpixels, points, or regions), which is useful for tasks such as segmentation, reconstruction, and clustering. In this work, we propose and evaluate a fully quantum pipeline for computing MSTs using the FALQON algorithm, a feedback-based quantum optimization method that does not require classical optimizers. We construct a Hamiltonian formulation whose ground-state energy encodes the MST of a graph and compare different FALQON strategies: (i) time rescaling (TR-FALQON) and (ii) multi-driver configurations. To avoid domain-specific biases, we adopt graphs with random weights and show that the FALQON variants exhibit significant differences in ground-state fidelity. We discuss the relevance of this approach for computer vision problems that naturally yield graph representations, and experimental results on synthetic instances together with a small demonstrative study on image segmentation illustrate both the potential and the current limitations of the method. Our numerical simulations on randomly weighted graphs show that standard one drive FALQON, although it reduces the expected energy, fails to concentrate amplitude in the MST solution. The multi drive variant succeeds in redistributing probability mass toward the ground state so that the MST appears among the most probable outcomes, and TR FALQON applied over multi drive produces the best results with faster convergence, lower final energy, and the highest solution state probability or fidelity in our tested instances. These improvements were observed on small synthetic graphs, underscoring both the promise of multi drive controls with temporal rescaling and the need for further scaling and hardware validation.}

\onecolumn \maketitle \normalsize \setcounter{footnote}{0} \vfill

\section{\uppercase{Introduction}}
\label{sec:introduction}

\begin{figure*}[t!]
    \centering
    \includegraphics[width=1\linewidth]{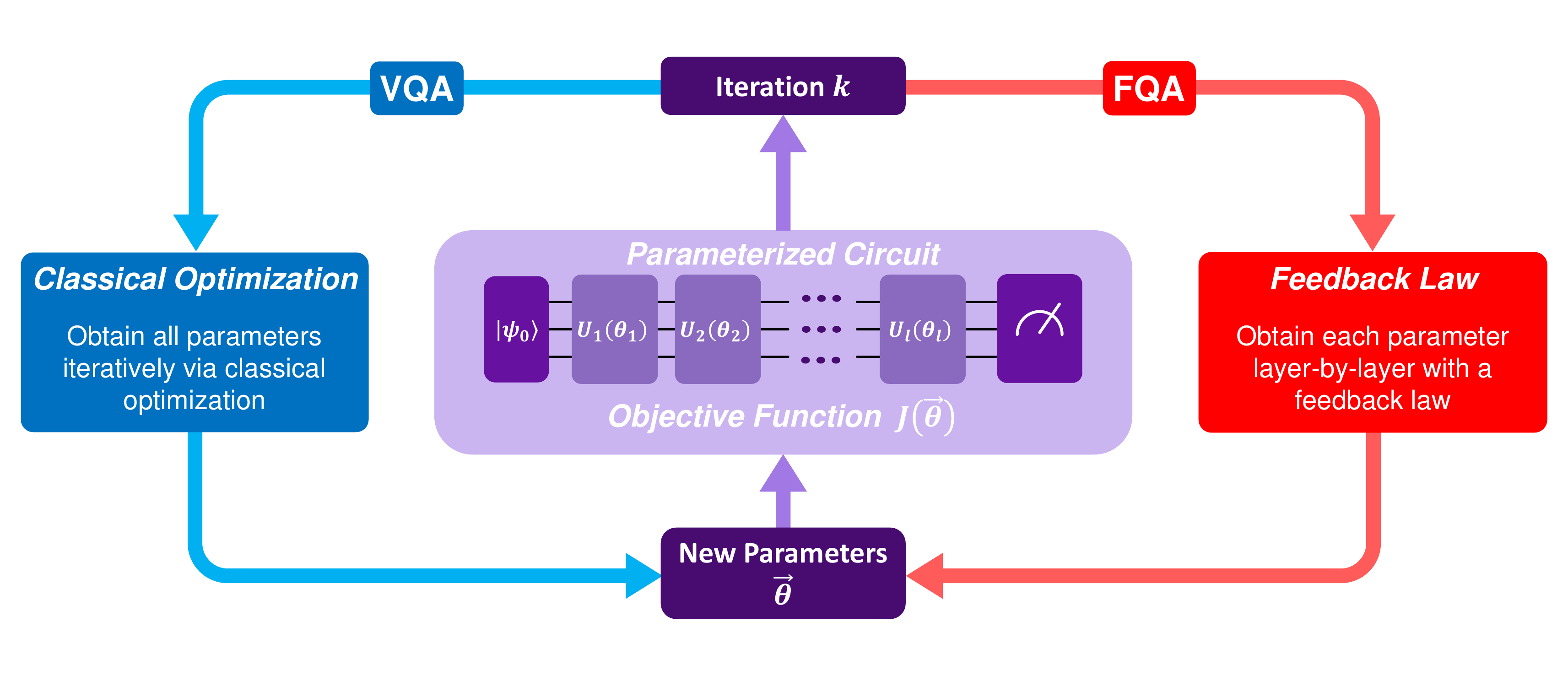}
    \caption{Diagram comparing VQAs and FQAs. Both can be formulated with the same objective function $J$ and use quantum circuits with an identical structure (center). The difference lies in how they minimize $J$: VQAs adjust all parameters $\vec{\theta}$ simultaneously through classical optimization (left), whereas FQAs eliminate this step and determine the parameters iteratively using a feedback law, layer by layer (right).}
    \label{fig:vqa}
\end{figure*}

Quantum computing has emerged as a promising framework for tackling combinatorial optimization problems whose complexity scales exponentially with system size. Recent advances in quantum algorithms and NISQ hardware have stimulated applications in engineering and computer science, including computer vision tasks that naturally rely on graph-based representations \cite{cheng2023noisy,mebtouche2024quantum}. In this context, the \emph{Minimum Spanning Tree} (MST) plays a central role in modeling similarity and connectivity relations among pixels, superpixels, or feature points, providing sparse and interpretable structures for segmentation, clustering, and surface reconstruction \cite{deRosa2014trainingRBF_OPF}.

While the MST can be solved efficiently by classical polynomial-time algorithms, practical vision pipelines often involve noise, additional constraints, or integration with probabilistic graphical models, motivating alternative optimization-based formulations. A common approach is to encode combinatorial problems as QUBO or Ising models, where the solution corresponds to the ground state of a problem Hamiltonian \cite{lewis2017qubo}. This enables the use of quantum algorithms for low-energy state preparation, including Variational Quantum Algorithms (VQAs) \cite{cerezo2021variational} and Feedback-based Quantum Algorithms (FQAs) \cite{magann2022feedback}.

A growing body of work reformulates vision problems such as segmentation, stereo matching, and medical image reconstruction as QUBO/Ising models, exploiting their underlying graph structure \cite{qseg2023,cruzsantos2018,heidari2024}. Related studies also propose QUBO formulations for spanning tree and network reconfiguration problems, where connectivity and acyclicity constraints are enforced via quadratic penalties \cite{carvalho2022}. These formulations are particularly relevant when images are represented as graphs of superpixels or features, allowing structural solutions such as MSTs to emerge from energy minimization.

In the NISQ regime, qubit-efficient encodings and hybrid pipelines have demonstrated the feasibility of applying quantum optimization to real-world vision problems, including medical imaging tasks \cite{jun2025,pge2024,Cieslak2018_seamCarving}. These results support the use of fully quantum approaches for graph-based vision problems.

\begin{figure*}[t!]
    \centering
    \includegraphics[width=1\linewidth]{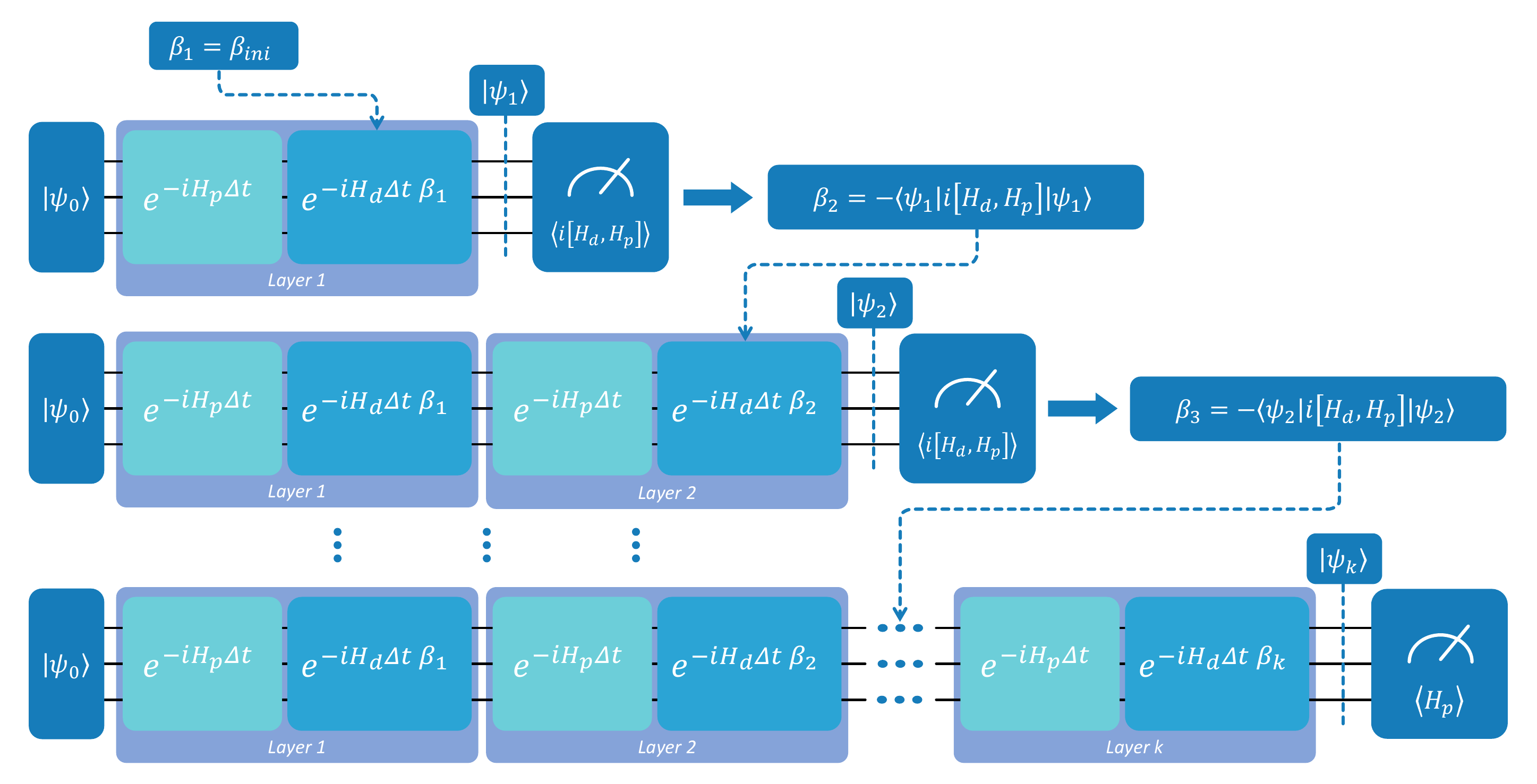}
    \caption{Illustrative diagram of the FALQON \cite{PhysRevB.110.224422}. The process begins with the state $\ket{\psi_0}$, and at each layer $k$, the unitary operators $e^{-iH_p \Delta t}$ and $e^{-iH_d \Delta t \beta_k}$ are applied sequentially. The parameter $\beta_k$ is adaptively adjusted at each iteration. This dynamic is repeated iteratively, guiding the evolution of the state $\ket{\psi}$ through the layers until the desired solution is reached.}
    \label{fig:diag_falqon}
\end{figure*}

In this work, we focus on FQAs, specifically the FALQON family, rather than hybrid variational methods. Unlike VQAs, which rely on an external classical optimizer and suffer from latency and optimization overhead, FQAs update parameters through local quantum feedback rules based on measurements performed on the system itself. This fully quantum training process reduces classical overhead and enables faster, layer-by-layer updates, albeit requiring real-time measurement and control capabilities.

Despite these advantages, systematic comparisons of FQA variants applied to QUBO formulations of vision-relevant problems remain limited. We introduce the FALQON-MST framework, which combines (i) a circuit-compatible QUBO/Ising formulation of the MST problem and (ii) a comparative study of three FALQON variants: standard FALQON, TR-FALQON with time rescaling, and multi-driver schemes. To avoid instance-specific bias, our experiments are conducted on randomly weighted graphs.

The main contributions of this work are:
\begin{itemize}
\item A fully quantum pipeline for solving the QUBO formulation of the MST problem with appropriate constraint encoding.
\item An implementation and comparison of standard, TR-FALQON, and multi-driver FALQON variants in terms of ground-state fidelity, layer efficiency, and robustness to measurement noise.
\item An experimental evaluation on synthetic graph instances highlighting the practical performance and limitations of FALQON-based approaches.
\end{itemize}

The remainder of the paper is organized as follows. Section~\ref{sec:MST} presents the QUBO/Ising formulation of the MST problem. Section~\ref{sec:FQA} reviews the FALQON algorithm and its variants. Section~\ref{sec:results} describes the experimental setup and results. Section~\ref{sec:conclusion} concludes with a discussion and future directions.

\section{\uppercase{Minimum Spanning Tree QUBO formulation}}
\label{sec:MST}

Given a weighted graph $G=(V,E,c)$, with vertices $V$, edges $E$, and edge costs $c(u,v)>0$, a minimum spanning tree (MST) is a spanning, acyclic subgraph $T=(V,E_T)$ that minimizes
\begin{equation}
C(T)=\sum_{(u,v)\in E_T} c(u,v).
\end{equation}
If each vertex is incident to at most $\Delta$ edges, the problem is referred to as a $\Delta$-MST.

Lucas \cite{lucas2014} introduced the first Ising/QUBO formulation of the MST, requiring $\mathcal{O}(|V|^3)$ variables and $\mathcal{O}(|V|^5)$ interactions, which makes it impractical for near-term quantum hardware with limited qubit counts. Fowler later improved this formulation \cite{fowler2017}, reducing the requirements to $\mathcal{O}(|V|^2)$ variables and $\mathcal{O}(|V|^3)$ interactions. This reformulation employs topological ordering and a fixed root node to enforce acyclicity and connectivity, bringing the MST problem to a complexity comparable to that of the Traveling Salesman Problem and enabling more feasible experimental implementations.

The formulation uses two sets of variables:
\begin{itemize}
     \item Edge Variables $(e_{u,v})$: Indicate whether an edge $(u,v) \in E$ is included in the minimum spanning tree:
    \begin{equation}
    e_{u,v} = 
    \begin{cases}
    1, & \text{if the edge $(u,v)$ is included}; \\
    0, & \text{otherwise}.
    \end{cases}
    \end{equation}
\end{itemize}

Concerning the root $v_0$, we just have $e_{v_0,u}$ but not $e_{u,v_0}$. All other edges go in both ways $e_{u,v}$ and $e_{v,u}$.

\begin{itemize}
    \item Order Variables $(x_{u,v})$: Represent a topological ordering of the vertices to ensure that the graph is acyclic:
    \begin{equation}
    x_{u,v} =
    \begin{cases}
    1, & \text{if vertex $u$ precedes vertex $v$;} \\
    0, & \text{otherwise}.
    \end{cases}
    \end{equation}

    The root does not take part in this, and only $x_{u,v}$ is included, not $x_{v,u}$.

\end{itemize}

The Hamiltonian consists of the following components:

\begin{itemize}
    \item Acyclicity: To enforce acyclicity, the term $F_{I,1}(x)$ penalizes configurations that violate a valid topological order:
    \begin{equation}
    \begin{aligned}
    F_{I,1}(x) &= \sum_{\substack{1 \le u < v < w \le |V|\\ u,v,w \neq v_0}}
    \Big( x_{u,w} + x_{u,v}x_{v,w} \\
    &\qquad - x_{u,v}x_{u,w} - x_{u,w}x_{v,w} \Big).
    \end{aligned}
    \end{equation}

    \item Edge-Order Consistency: This term aligns edge inclusion with the vertex ordering:
    \begin{equation}
    F_{I,2}(e,x)
    \;=\;
    \sum_{\substack{(u,v)\in E\\ u < v,\, u,v \neq v_0}}
    \big( e_{u,v}(1 - x_{u,v}) + e_{v,u}\,x_{u,v} \big).
    \end{equation}

    \item Connectivity: To ensure that all vertices (except the root $v_0$) are connected, $F_{I,3}(e)$ penalizes vertices without incoming edges:
    \begin{equation}
    F_{I,3}(e)
    \;=\;
    \sum_{v \in V \setminus \{ v_0 \}} \left( 1 - \sum_{(u,v)\in E} e_{u,v} \right)^2.
    \end{equation}

    \item Cost Function: This term minimizes the total weight of the spanning tree:
    \begin{align}
    O_I(e)
    &= 
    \sum_{\substack{(u,v)\in E\\ u,v \neq v_0}} 
        c(u,v)\,\big( e_{u,v} + e_{v,u} \big)
    \nonumber\\[4pt]
    &\quad+\;
    \sum_{(v_0,u)\in E} 
        c(v_0,u)\, e_{v_0,u}.
    \end{align}
\end{itemize}

\begin{itemize}
    \item Complete Hamiltonian: The full Hamiltonian combines these components with a penalty coefficient $P_I$ to prioritize constraint satisfaction:
    \begin{equation}
    \begin{split}
    H_I(e,x) &= P_I\big( F_{I,1}(x) + F_{I,2}(e,x) \\
    &\qquad + F_{I,3}(e) \big) + O_I(e).
    \end{split}
    \end{equation}
    
    where
    \begin{equation}
    P_I \;=\; (|V| - 1)\cdot m + 1,
    \end{equation}
    and
    \begin{equation}
    m \;=\; \max_{(u,v)\in E} c(u,v).
    \end{equation}
\end{itemize}

\begin{figure*}[t!]
    \centering
    \includegraphics[width=1\linewidth]{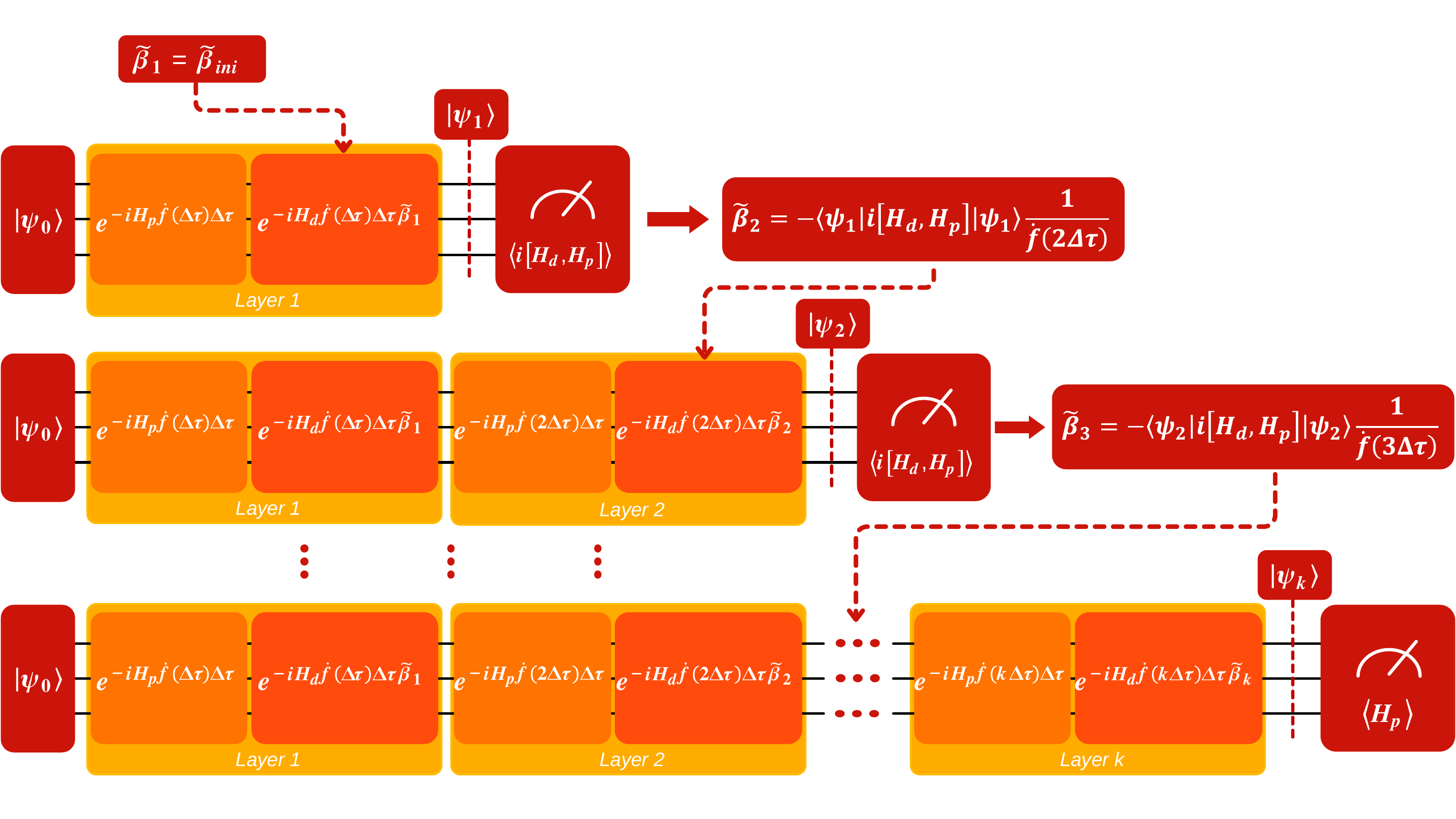}
    \caption{Illustrative diagram of the TR-FALQON \cite{qc91-5mj2}. The process starts with the initial state $\ket{\psi_0}$. In each layer $k$, the unitary operators $e^{-iH_p \dot{f}(k\Delta \tau) \Delta \tau}$ and $e^{-iH_d \dot{f}(k\Delta \tau) \Delta \tau \tilde{\beta}_k}$ are applied sequentially, adaptively adjusting the parameter $\tilde{\beta}_k$.}
    \label{fig:rtfqa}
\end{figure*}

\section{\uppercase{feedback-based quantum algorithms}}
\label{sec:FQA}

\subsection{FALQON}

The method is based on Lyapunov control applied to the Hamiltonian $H(t)=H_p+\beta(t)H_d$:
\begin{equation}
i\frac{d}{dt}\ket{\psi(t)}=(H_p+\beta(t)H_d)\ket{\psi(t)}.
\end{equation}
We define the cost functional $J(t)=\bra{\psi(t)}H_p\ket{\psi(t)}$ and
\begin{equation}
A(t)=\bra{\psi(t)} i[H_d,H_p]\ket{\psi(t)},
\end{equation}
such that
\begin{equation}
\frac{dJ}{dt}=A(t)\beta(t).
\end{equation}
Choosing the control law
\begin{equation}
\beta(t)=-A(t),
\end{equation}
we have $\dot J=-A(t)^2\le0$, guaranteeing monotonic decay of the cost (provided $[H_d,H_p]\neq 0$).

In the trotterized implementation (layered circuit) the discrete evolution is
\begin{equation}
\ket{\psi_k}=U_d(\beta_k)U_p\cdots U_d(\beta_1)U_p\ket{\psi_0},
\end{equation}
with $U_p=e^{-iH_p\Delta t}$ and $U_d(\beta_k)=e^{-i\beta_k H_d\Delta t}$.  
The discrete feedback law is simply
\begin{equation}
\beta_k=-A_{k-1},\qquad A_{k-1}=\bra{\psi_{k-1}} i[H_d,H_p]\ket{\psi_{k-1}}.
\end{equation}

\subsection{TR-FALQON}

The idea is to apply a time rescaling $t=f(\tau)$ to obtain a TR Hamiltonian
\begin{equation}
H_{\text{TR}}(\tau)=H(f(\tau))\dot f(\tau),
\end{equation}
so that the same final evolution is achieved on an interval $\tau\in[0,\tau_f=f^{-1}(t_f)]$ (possibly shorter). For the TR to serve as a convenient adiabatic shortcut we choose $f$ such that
\[
f^{-1}(0)=0,\quad f^{-1}(t_f)<t_f,\quad f'(0)=f'(f^{-1}(t_f))=1,
\]
for example
\begin{equation}
f(\tau)=a\tau-\frac{t_f}{2\pi a}(a-1)\sin\!\left(\frac{2\pi a\tau}{t_f}\right),\quad a>1.
\end{equation}

Applying the same Lyapunov logic in the variable $\tau$, we define $A(\tau)=\bra{\psi(\tau)} i[H_d,H_p]\ket{\psi(\tau)}$ and choose
\begin{equation}
\tilde\beta(\tau)=-\frac{A(\tau)}{\dot f(\tau)},
\end{equation}
which ensures $\dfrac{d}{d\tau}\bra{\psi}H_p\ket{\psi}\le0$ (using $H_{\text{TR}}$).

In the trotterized discrete form (step $\Delta\tau$) the TR evolution is
\begin{equation}
\ket{\psi_k}=\tilde U_d(\tilde\beta_k)\tilde U_p\cdots \tilde U_d(\tilde\beta_1)\tilde U_p\ket{\psi_0},
\end{equation}
with
\[
\begin{aligned}
\tilde U_p 
&= \exp\!\big(-iH_p\,\dot f(k\Delta\tau)\,\Delta\tau\big), \\
\tilde U_d(\tilde\beta_k) 
&= \exp\!\big(-iH_d\,\tilde\beta_k\,\dot f(k\Delta\tau)\,\Delta\tau\big).
\end{aligned}
\]
and the practical feedback law
\begin{equation}
\tilde\beta_k=-\frac{A_{k-1}}{\dot f(k\Delta\tau)},
\end{equation}
where $A_{k-1}=\bra{\psi_{k-1}}i[H_d,H_p]\ket{\psi_{k-1}}$.

\begin{figure*}[t!]
\centering
\includegraphics[width=0.8\linewidth]{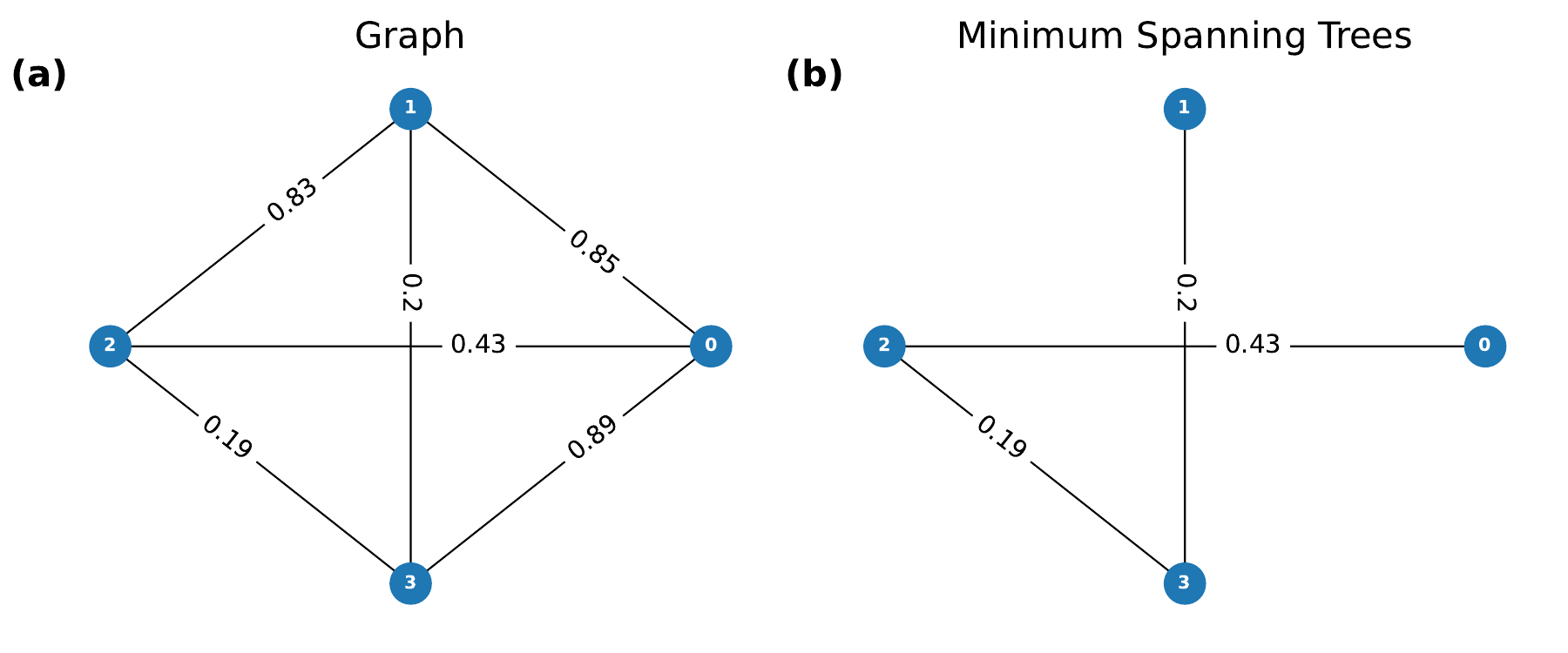}
\caption{(a) Example of a graph constructed with randomly generated edge weights. (b) Minimum Spanning Tree derived from the graph in (a), where the highest-cost edges are removed while preserving connectivity among the nodes at minimal total cost.}
\label{fig:graph_mst}
\end{figure*}

\subsection{Multi Drive}

The structure of Quantum Lyapunov Control and FALQON can be directly extended to configurations with multiple control functions \cite{PhysRevA.106.062414}, that is, when the system Hamiltonian is given by
\begin{equation}
H(t) = H_p + \sum_j \beta(j,t) H_{d,j},
\label{eq:multiH}
\end{equation}
where $\beta(j,t)$ denotes the value of the control function that scales the $j$-th driving Hamiltonian $H_{d,j}$ at time $t$.
Then, to satisfy the QLC condition that $\frac{d}{dt}E_p \leq 0$, we have to
\begin{align}
\frac{d}{dt}\langle \psi(t)|H_p|\psi(t)\rangle 
&= \langle \psi(t)|[H_p + \sum_j \beta(j,t)H_{d,j}, H_p]|\psi(t)\rangle \notag\\
&= \langle \psi(t)|[\sum_j \beta(j,t)H_{d,j}, H_p]|\psi(t)\rangle \notag\\
&= \sum_j \langle \psi(t)|[H_{d,j}, H_p]|\psi(t)\rangle \beta(j,t) \notag\\
&= \sum_j A(j,t)\beta(j,t),
\label{eq:dEp_multi}
\end{align}
where
\begin{equation}
A(j,t) = \langle \psi(t)|[H_{d,j}, H_p]|\psi(t)\rangle.
\label{eq:Adef}
\end{equation}

Thus, the following control laws can be used:
\begin{equation}
\beta(j,t) = -w_j f_k(A(j,t)), \quad \forall j,
\label{eq:control_law}
\end{equation}
to ensure that $E_p$ decreases monotonically over time.

We apply the approach described above to FALQON and refer to it as Multi Drive.

\begin{figure*}[t!]
    \centering
    \includegraphics[width=0.84\linewidth]{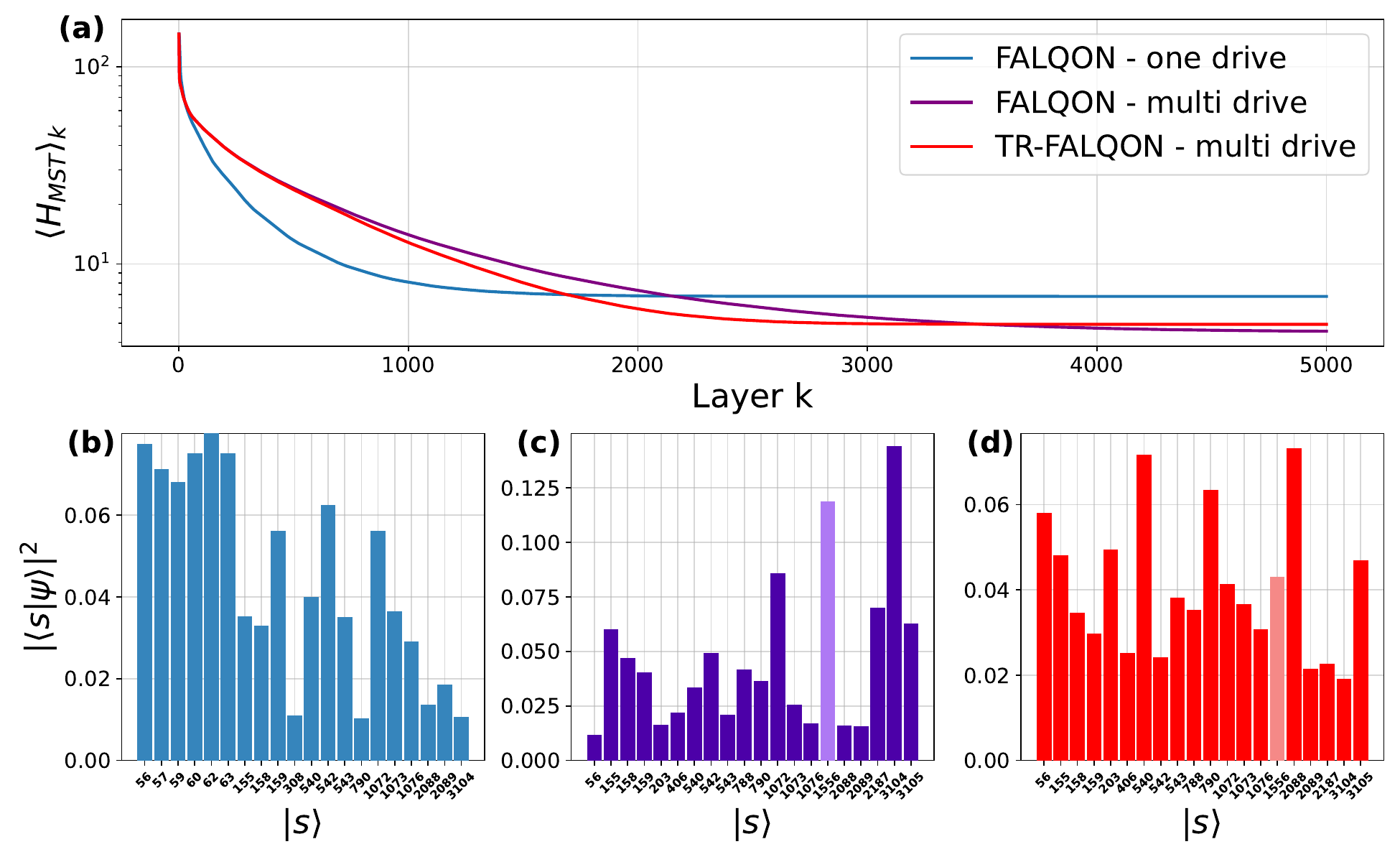}
    \caption{Numerical simulation results for the MST Hamiltonian formulated as a QUBO with time step $\Delta t=0.02$. (a) Convergence of the cost $J=\langle\Psi_k|H_{MST}|\Psi_k\rangle$ as a function of the number of layers $k$ for FALQON (one-drive), FALQON (multi-drive), and TR-FALQON (multi-drive); energy decreases with the number of layers, with TR-FALQON multi-drive achieving the lowest final energy and fastest convergence. (b) Probability distribution of basis states at the end of the protocol for FALQON one-drive, where the state encoding the MST does not appear among the most probable. (c) Analogous distribution for FALQON multi-drive, where the solution state appears with significant probability, indicating amplitude concentration in the ground state. (d) TR-FALQON multi-drive, showing higher probability concentration in the solution state and a less dispersed distribution, consistent with the lower final energy shown in (a).}
    \label{fig:resultado}
\end{figure*}

\section{\uppercase{Results}}
\label{sec:results}

In this section, we show how a quantum computer can be used to study problems in computer vision encoded in graphs and MSTs using FALQON and TR-FALQON, as well as their Multi-Drive versions.

Figure \ref{fig:resultado} shows numerical simulations applied to the MST-QUBO Hamiltonian depicted in Figure \ref{fig:graph_mst}. The simulations use $\Delta t = 0.02$. Panel (a) of Figure \ref{fig:resultado} illustrates the convergence of the cost function, $J = \langle \Psi_k | H_{MST} | \Psi_k \rangle$, as a function of the number of layers. The expectation value of the objective function decreases monotonically with the number of layers, demonstrating the convergence of FALQON and TR-FALQON and their Multi-Drive versions to the ground state. Panels (b), (c), and (d) of Figure \ref{fig:resultado} show the probabilities of the most likely states for FALQON one-drive, FALQON multi-drive, and TR-FALQON multi-drive, respectively; the state encoding the MST appears only in the multi-drive variants (panels (c) and (d)), indicating that standard FALQON and, empirically, TR-FALQON in single-drive do not reach the correct solution, highlighting the necessity of the multi-drive scheme.

The numerical simulations were performed on the MST Hamiltonian formulated as a QUBO, with randomly generated edge weights to avoid instance bias (see Figure \ref{fig:graph_mst}). Figure \ref{fig:resultado} summarizes the comparative results obtained for the FALQON algorithm variants used in this work: standard FALQON (one-drive), FALQON with multi-drive, and TR-FALQON with multi-drive.

Panel (a) of Figure \ref{fig:resultado} presents the convergence curve of the expected value of the problem Hamiltonian as a function of the number of layers/iterations k. It is observed that standard one-drive FALQON experiences an initial energy reduction, followed by a plateau that prevents reaching the minimum energy associated with the correct solution. The FALQON multi-drive variant reduces the energy over more layers and achieves a lower final value than the one-drive version, indicating a greater ability to escape local minima. The TR-FALQON with multi-drive combination demonstrates the best performance, with faster convergence and lower final energy, suggesting that temporal rescaling acts as an effective adiabatic shortcut when applied over a multi-drive scheme.

Panel (b) of Figure \ref{fig:resultado} shows the probabilities of the most likely basis states measured at the end of the protocol for standard FALQON (one-drive). In this case, the correct state encoding the minimum spanning tree does not appear among the most probable states. The probability mass is spread across several incorrect configurations, indicating that mere reduction of the average energy does not guarantee amplitude concentration in the solution state.

Panel (c) of Figure \ref{fig:resultado} presents the analogous probability distribution for FALQON multi-drive. Here, the correct state appears among the states with relevant probability, showing that multi-drive tends to concentrate a non-negligible fraction of amplitude in the target state. Compared to panel (b), there is a clear redistribution of probability mass towards lower-energy configurations, reflecting increased fidelity to the ground state.

Panel (d) of Figure \ref{fig:resultado} corresponds to TR-FALQON with multi-drive. In this case, the correct state appears with even higher probability than in panel (c) and often constitutes the most probable state at the end of the protocol. The probability distribution is less dispersed, consistent with the lower final energy and accelerated convergence observed in panel (a). These results suggest that temporal rescaling enhances the benefits of multi-drive, favoring evolution trajectories that concentrate amplitude in the ground state.

Taken together, the four panels indicate that standard one-drive FALQON cannot, in the tested instances, concentrate amplitude in the state encoding the MST solution, even when the expected Hamiltonian value decreases. The introduction of multi-drivers allows the algorithm to explore additional control directions whose contribution transforms energy reduction into an effective increase in the probability of the correct state. Temporal rescaling (TR) further improves this concentration when applied over the multi-drive scheme, acting as a shortcut that preserves the trajectory leading to the ground state. Although we do not present a specific panel for TR-FALQON in single-drive configuration, empirical observations indicate behavior similar to standard FALQON, highlighting the necessity of multi-drive to reach the correct solution in the considered instances.

\section{\uppercase{Conclusions}}
\label{sec:conclusion}

In this work, we present FALQON-MST, a fully quantum pipeline for solving the QUBO formulation of the Minimum Spanning Tree (MST) problem, and we systematically compare three variants based on quantum feedback laws: standard FALQON (one-drive), FALQON with multiple drivers (multi-drive), and TR-FALQON (temporal rescaling) applied over the multi-drive scheme. Simulations on instances with random weights indicate that mere reduction of the average energy does not always lead to amplitude concentration in the state encoding the solution; the introduction of multiple drivers favors probability redistribution toward the ground state; and temporal rescaling enhances these effects, accelerating convergence and increasing fidelity to the solution state in the tested instances.

From a computer vision perspective, these results are relevant both conceptually and practically. Graphs naturally arise in image representations, for example, superpixel graphs, 3D point clouds, and adjacency graphs between regions; an MST provides a sparse and interpretable representation of connectivity that can be used as a skeleton for hierarchical segmentation, boundary extraction, reconstruction, and clustering. The demonstration that quantum feedback-based blocks can concentrate amplitude in the solution state suggests that such quantum primitives could be integrated into vision pipelines to build or optimize graph structures that subsequently feed classical processing or classification methods.

We acknowledge important limitations: the experiments reported here are small-scale and use synthetic instances; the QUBO formulation requires careful penalty definition; and execution on NISQ hardware still faces practical constraints, such as the number of qubits, noise, and gate fidelity. We do not claim a universal advantage over the best classical MST algorithms; our goal is to explore how adaptive quantum controls can complement vision workflows, particularly in scenarios enriched by additional constraints, measurement uncertainties, or connectivity optimization requirements within hybrid pipelines.

For future work, we propose the following concrete directions:

\begin{itemize}
\item Scaling and validation on real data, assess the robustness of the variants such as one-drive, multi-drive and TR on larger graphs and real image datasets, measuring impacts on vision metrics such as IoU, F-score, and ARI, as well as quantum metrics such as fidelity and solution state probability.
\item Hybrid integration: study strategies in which the quantum MST serves as initialization or constraint for classical methods, for example, for graph pruning before segmentation or for constructing sparse neighborhoods used by classifiers.
\item Application to the supervised Optimum-Path Forest (OPF) algorithm, investigate two complementary approaches, namely (i) formulate OPF components, such as seed selection or path costs, as Hamiltonians whose ground state represents desired configurations and apply FALQON/TR-FALQON to optimize them; and (ii) use the MST produced by quantum methods as the underlying graph structure for OPF, evaluating gains in accuracy and efficiency.
\item Implementation on noisy hardware, port the protocols to noisy simulators and, when possible, test on available quantum backends, evaluating error sensitivity, measurement cost, and mitigation techniques.
\item Parameter sensitivity analysis, study performance dependence on choices of $\Delta t$, penalties, and rescaling functions $f(\tau)$, and develop automatic heuristics for practical configuration in vision problems.
\end{itemize}

In summary, FALQON-MST indicates that quantum feedback-based controls, particularly the combination of multi-drive and temporal rescaling, are promising as building blocks for structural tasks in computer vision. The results motivate further experimental and theoretical investigations to transform these adaptive quantum blocks into functional and integrable components within hybrid vision pipelines.

\section*{\uppercase{Acknowledgements}}

The authors are grateful to the São Paulo Research Foundation (FAPESP) grants 2024/18453-6, 2019/07665-4, 2021/04655-8, 2023/03726-4, 2023/10823-6, 2023/12830-0, 2023/14427-8, 2025/13172-1, and to the Brazilian National Council for Scientific and Technological Development (CNPq) grant 308529/2021-9, to the Petrobras grant 2023/00466-1, and to the Office of Naval Research (ONR) grant N62909-24-1-2012.

\bibliographystyle{apalike}
{\small
\bibliography{example}}

\end{document}